# Rapid characterization of wafer-scale 2D material: Epitaxial graphene and graphene nanoribbons on SiC

Vishal Panchal[*1,2], Yanfei Yang[*1,3], Guangjun Cheng[1], Jiuning Hu[1], Chieh-I Liu[1,4], Albert F. Rigosi[1], Christos Melios[2,5], Olga Kazakova[2], Angela R. Hight Walker[1], David B. Newell[1], and Randolph E. Elmquist[*1]

[1] National Institute of Standards and Technology, Gaithersburg, MD 20899, USA

[2] National Physical Laboratory, Hampton Road, Teddington, TW11 0LW, UK

[3] Joint Quantum Institute, University of Maryland, College Park, MD 20742, USA

[4] National Taiwan University, Taipei, 10617, Taiwan

[5] Advanced Technology Institute, University of Surrey, Guildford, Surrey, GU2 7XH, UK

**ABSTRACT:** We demonstrate that the confocal laser scanning microscopy (CLSM) provides a non-destructive, highly-efficient characterization method for large-area epitaxial graphene and graphene nanostructures on SiC substrates, which can be applied in ambient air without sample preparation and is insusceptible to surface charging or surface contamination. Based on the variation of reflected intensity from regions covered by interfacial layer, single layer, bilayer, or few layer graphene, and through the correlation to the results from Raman spectroscopy and SPM, CLSM images with a high resolution ($\approx 150$ nm) reveal that the intensity contrast has distinct feature for undergrown graphene (mixing of dense, parallel graphene nanoribbons and interfacial layer), continuous graphene, and overgrown graphene. Moreover, CLSM has areal acquisition time hundreds of times faster per unit area than the supplementary characterization methods. We believe that the confocal laser scanning microscope will be an indispensable tool for mass-produced epitaxial graphene or applicable 2D materials.

Wafer-scale graphene material is of strong interest for quantum Hall resistance standards[1-5] and future nanoelectronics[6,7], such as high frequency electronics[8-15] and photonics[16,17]. The high uniformity of single-domain epitaxial graphene (EG) grown on the silicon face of SiC(0001)[18] presents several advantages, such as the lack of a need to transfer the graphene onto an insulating substrate for device processing, as is the case with chemical vapor deposition (CVD) growth. Recent progress[5,19,20] in EG has demonstrated the potential for mass production of homogeneous EG at the wafer-scale, naturally leading to a demand for a characterization method that is fast, accurate and accessible. Currently, Raman spectroscopy and scanning probe microscopies (SPM) including atomic force microscopy (AFM) and Kelvin probe force microscopy (KPFM) are the most widely used methods of characterizing EG quality. Raman spectroscopy is a fast and non-destructive tool to identify monolayer graphene, whose fingerprint in the Raman spectrum is a symmetric G' peak (2D band) at ~2700 cm$^{-1}$ that can be fitted by a single Lorentzian[21]. While the shape of G' peak evolves quickly with increased number of layer, careful analysis of G'(2D) band is required to identify bilayer and thicker graphene. Moreover, Raman spectra are sensitive to the doping level and strain in EG[22-28]. Topography imaged by AFM shows the SiC terrace morphology, which develops concurrently with the EG and thus has a strong influence on the layer growth and uniformity[5,29]. The identification of EG domains from their thickness is far from straight-forward using AFM alone[30]. Recently, KPFM has been shown to be a reliable method for distinguishing the layer number up to several layers[31]. However, Raman and SPM methods are time consuming and are suitable only for sampling in nanoelectronics applications over regions of tens of micrometers. More universal electronics applications require fast and accurate characterization of graphene structures at the wafer scale while at the same time retaining sub-micrometer scale spatial resolution. Prior to our work, transmitted optical microscopy has been demonstrated by Yager *et al.*[32] for rapid identification of layer-number inhomogeneity in graphene over hundred-micrometer scales.

In this report, we demonstrate that confocal laser scanning microscopy (CLSM) using reflected optical light is a superior tool for characterization of large-area epitaxial graphene and graphene nanostructures grown on SiC, compared to conventional optical microscope, Raman spectroscopy, AFM, conductive atomic force microscopy (C-AFM), KPFM, and scanning electron microscope (SEM) methods. CLSM has a lateral resolution that is pushed beyond the optical diffraction limit and depth-of-field is improved with stacking of images at different focal planes, allowing high resolution over larger imaging areas. We first describe the experiment setup of CLSM, SPM, and Raman spectroscopy. Then we discuss the CLSM imaging results, which show the distinct intensity contrast for undergrown, continuous, and overgrown EG, through the correlation to Raman spectroscopy and SPM.

**RESULTS**

Samples for this study were processed in a graphite-lined resistive-element furnace in Ar background near atmospheric pressure[3]. Graphene can be grown with close to monolayer thickness on the Si-face of semi-insulating 4H-SiC substrates with the SiC(0001) side facing down toward a polished glassy graphite disk. This method has been often referred as FTG (facing-to-graphite) growth in our previous publications[3-5]. Before being loaded into the furnace, the SiC substrate is immersed in buffered HF (< 10 %) for one minute. We adopt a two-stage annealing process, which pre-treats the samples at 1050 $^\circ$C in forming gas (4 % $H_2$ + 96 % Ar) for two hours, and then grows graphene at 1900 $^\circ$C in high purity Ar for less than 10 minutes.

**1.   Epitaxial graphene nanoribbons**

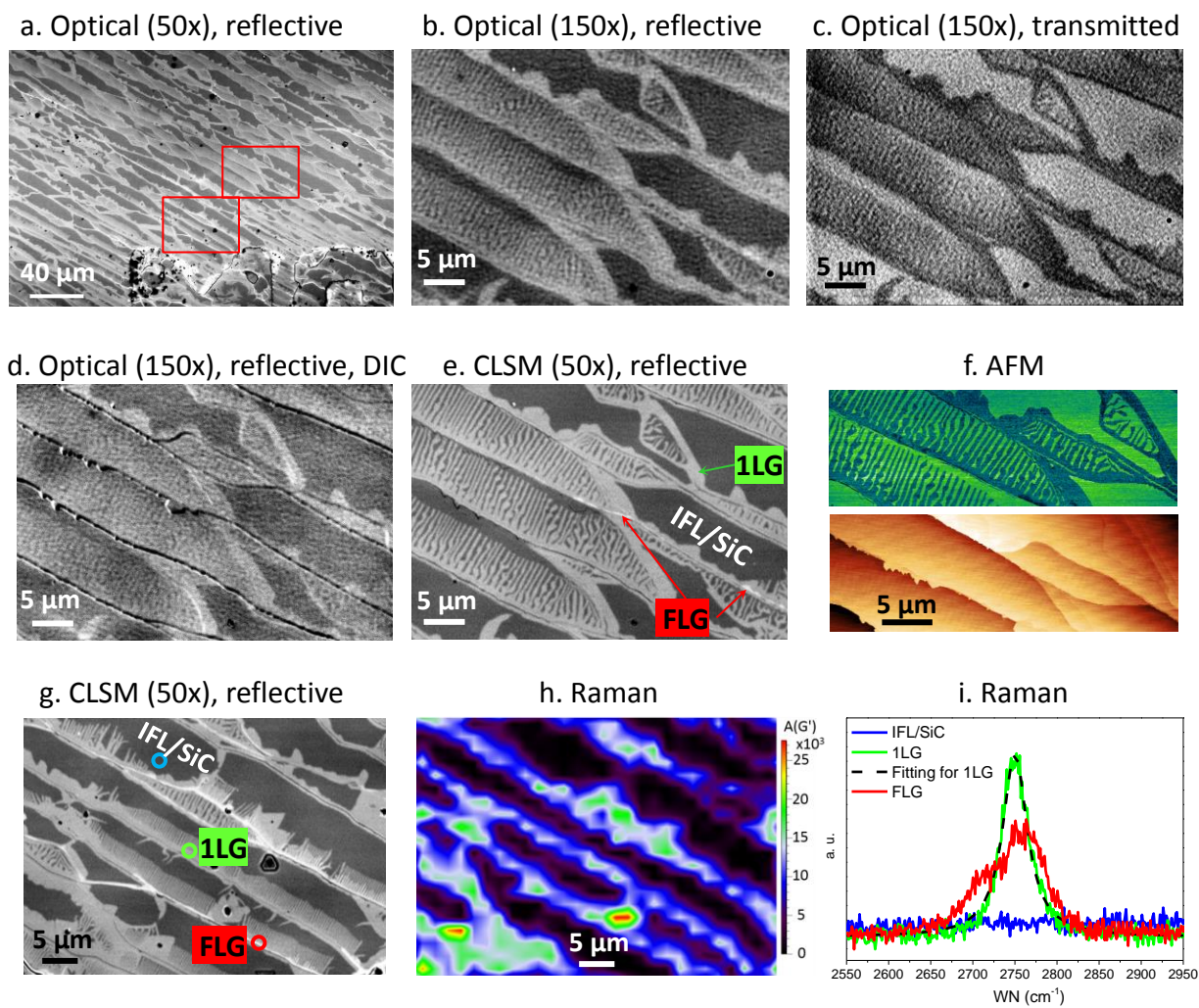

**Figure 1.** Epitaxial graphene nanoribbons on SiC. (a) Optical image taken with reflective illumination and a 50× objective. (b) Optical image of the area in the upper red box in (a) taken with reflective illumination using a 150× objective. (c) Optical image of the same area as in (b) taken with transmitted light and a 150× objective. (d) Reflective DIC image of the same area as in (b) taken with a 150× objective. (e) CLSM image of the same area as in (b) taken with a 50x objective. (f) AFM phase (top panel) and topography (bottom panel) images of the upper portion of (b). (g) CLSM image of the same area as in the lower red box in (a). (h) Mapping of the integrated G'(2D) peak area in the Raman spectra of the same area as in (g). (i) G'(2D) peaks of point A (IFL or SiC), point B (graphene ribbons), and point C (step edge) in (g).

The first four optical images in Figure 1 (a-d) are obtained by a conventional Nikon Eclipse L200N optical microscope [see Notes] fitted with both reflective and transmitted white light illumination, as well as a differential interference contrast (DIC) slider. Figure 1a shows the optical image of a large area (288 µm × 192 µm) on a sample with partial coverage of graphene, obtained with reflective illumination. Part of the fiducial mark "M0" that is etched ≈ 600 nm deep in the SiC substrate can be seen at the bottom. Figure 1b is the magnified reflective image of the area indicated by the upper red box in Fig. 1a, which reveals that the brighter regions contain a dense 2D forest of nanoribbons that are not fully resolved. Note that Fig. 1b (as well as Figs. 1c,1d,1f,1g) is a cropped region for the convenience of comparison with other corresponding images in Fig. 1.

Figure 1c shows the image of the same area as Fig. 1b obtained with transmitted illumination. The darker regions in Fig. 1c indicate the formation of monolayer graphene, which absorbs approximately 2.3% of the incident light[33]. The brighter regions are insulating IFL or bare SiC, as verified later with C-AFM in Fig. 2. With the DIC technique, we obtain a 3D vision of the terraces on the SiC surface formed during the high temperature annealing (Fig. 1d), with a trade-off for the nanoribbon contrast. From Fig. 1b-1d, we can see that the nanoribbons preferentially grow from the step edge towards the adjacent upper terrace. The formation of parallel nanoribbons on SiC(0001) terraces has been attributed to diffusion-limited growth[34] that may accompany the decomposition of single SiC atomic layers.

Fig. 1e is the CLSM image of the same area as in Fig. 1b, which possesses greatly enhanced lateral resolution and contrast through the point illumination from a laser source and by removal of the out-of-focus background light with a pinhole at the conjugate focus plane in front of the sensor. Compared to the conventional optical image in Fig. 1b, the CLSM image not only shows a much higher special resolution, but also clearly reveals thin stripes (labeled by red arrows) and patches of higher reflectivity along the step edges, indicating FLG. While the AFM phase image in Fig. 1f (upper panel) reveals comparable and higher resolution for the 2D nanoribbon forest, it mostly fails to distinguish the thin stripes and patches of thicker graphene layers along the step edge (neither feature can be distinguished from the AFM height image in the lower panel). Moreover, the acquisition time for a CLSM image (only 10 seconds for Fig. 1f using manual mode), is hundreds of times faster than that for a typical AFM image (15 minutes for Fig. 1e).

Raman spectra for the same region as in the lower red box in Fig. 1a were collected (in about 1.5 hours) to create a map of the integrated G'(2D) peak (Fig. 1h), that clearly shows the area that is covered by graphene. However, due to the relatively large laser spot size ($\approx 1$ μm), it cannot distinguish individual nanoribbons. Representative Raman spectra from points A, B and C, indicated by the red circles in Fig. 1g, are shown in Fig. 1i. There is no G'(2D) signal for the dark region (IFL/SiC) indicated by point A. The G'(2D) spectrum at point B shows a symmetric peak centered around 2750 $cm^{-1}$, which can be fitted by a single Lorentzian (dashed blue line in Fig. 1i) with a FWHM of $\approx 35$ $cm^{-1}$, indicating that the ribbons are monolayer graphene[5,35]. The G'(2D) peak of point C, which has a higher brightness than the ribbons, is significantly broadened with a clear shoulder (red line in Fig. 1i), indicative of multilayer or strained graphene[35,36].

Figure 2a is a map of the current channel obtained by C-AFM on a similar sample as characterized in Fig. 1, with incomplete coverage of graphene. The CLSM image (Fig. 2b) of the same region shows comparable resolution as the current map in Fig. 2a. To remove contamination that interferes with C-AFM, we applied mechanical cleaning over the graphene surface by scanning the sample with a stiff (~3 $Nm^{-1}$) AFM probe in contact mode. During the cleaning process, some of the graphene nanoribbons had been damaged and thus were electrically isolated from the rest of the network of graphene nanoribbons. These nanoribbons showed a distinct lack of current in the C-AFM map (see top right box in Fig 2a), but existed in the CLSM image of the same area taken ***after*** the cleaning process (see top right box in Fig. 2b). Obviously, CLSM has the advantage of being a non-invasive method for the characterization of the graphene ribbons on a SiC substrate. Furthermore, the current variation (color variation) in Fig. 1a does not represent different conductivity, but is due to the differences in the connection paths to the conductive AFM tip. As the result, while C-AFM confirms the conductivity of the graphene ribbons, as shown in Fig. 2a, it could not distinguish between the single-layer graphene nanoribbons and the multilayer graphene along the step edges, which is clearly revealed by CLSM imaging.

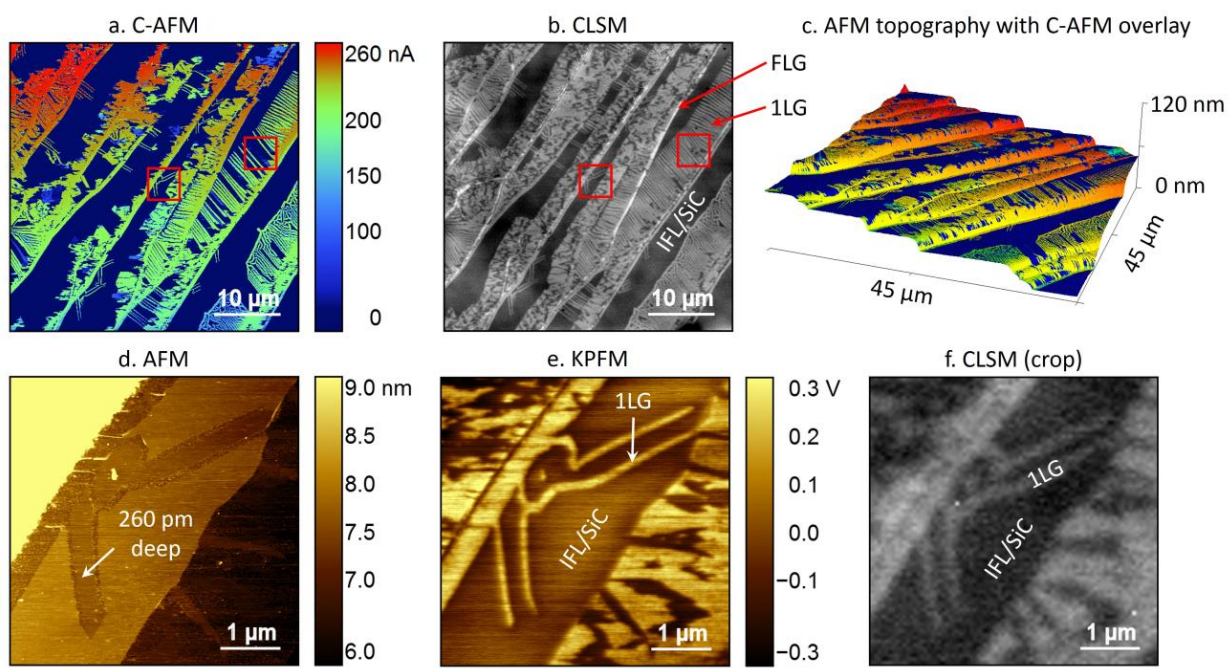

**Figure 2.** Single-layer carbon nanoribbon forest and multilayer carbon nanoribbons on SiC. (a) Map of conductivity of a region with incomplete coverage of graphene obtained by C-AFM (45 μm by 45 μm). (b) CLSM image of the same area as in (a). (c) 3D-display of the same area created by overlapping the current map and the height map obtained by C-AFM. (d)-(f) AFM topography (d), surface potential (e), and CLSM images (f) of the area indicated by the box near the center of the image in (a).

Figure 2c is a 3D image produced by overlapping the current channel with the height channel from the C-AFM scanning, again showing that the majority of graphene nanoribbons grow from the step edge to the upper steps. High-resolution AFM scanning reveals that the graphene nanoribbons are about 260 pm below the terrace surface (Fig. 2d), close to the thickness of one layer of hexagonal SiC. Figure 2e and 2f are surface potential and CLSM images, respectively, of the small region within the box near the center of the image in Fig. 2b, showing graphene nanoribbons of widths ≈100 nm.

The parallel graphene nanoribbons shown above are observed in EG samples produced by FTG growth (Fig. S1a), when the process temperature is not high enough or the process time is not long enough to produce full graphene coverage. Therefore, the dense 2D graphene nanoribbon forest along with its conspicuous optical contrast to the IFL patches is a very strong evidence of incomplete EG coverage and can be used alone for quick material assessment. The graphene nanoribbons will merge to form continuous graphene in a succeeding growth (Fig. S1), and the CLSM contrast features will evolve accordingly.

## 2. Overgrown epitaxial graphene

Next, we investigate EG samples with full coverage of graphene by comparing CLSM to KPFM results. The full coverage of graphene was produced by growth for ≈207 s at 1900 °C. Figure 3a is the CLSM intensity image showing a sample region predominantly covered by 1LG and having roughly 10 % coverage of 2LG and 3 LG domains as narrow patches (bright contrast in Figure 3a). Figure 3b is the CLSM image of the 15 μm by 15 μm area indicated by the red box in Fig. 3a, taken with 8x digital zoom. Comparing Fig. 3b to the surface potential map (Fig. 3c) of the same region confirms the designation of 1LG, 2LG, and 3LG. The higher reflective intensity from thicker graphene layer in the CLSM image is consistent with the linear increase of reflectivity reported by Ivanov et al. through monitoring the power of the reflected laser beam from EG samples on SiC[37]. Some dark lines and patches in Fig. 3b indicated by the red arrows are missing in the KPFM image due to the proximity effect of the charge potential, but are confirmed to be insulating buffer layer by the C-AFM, as shown in Fig. 3d.

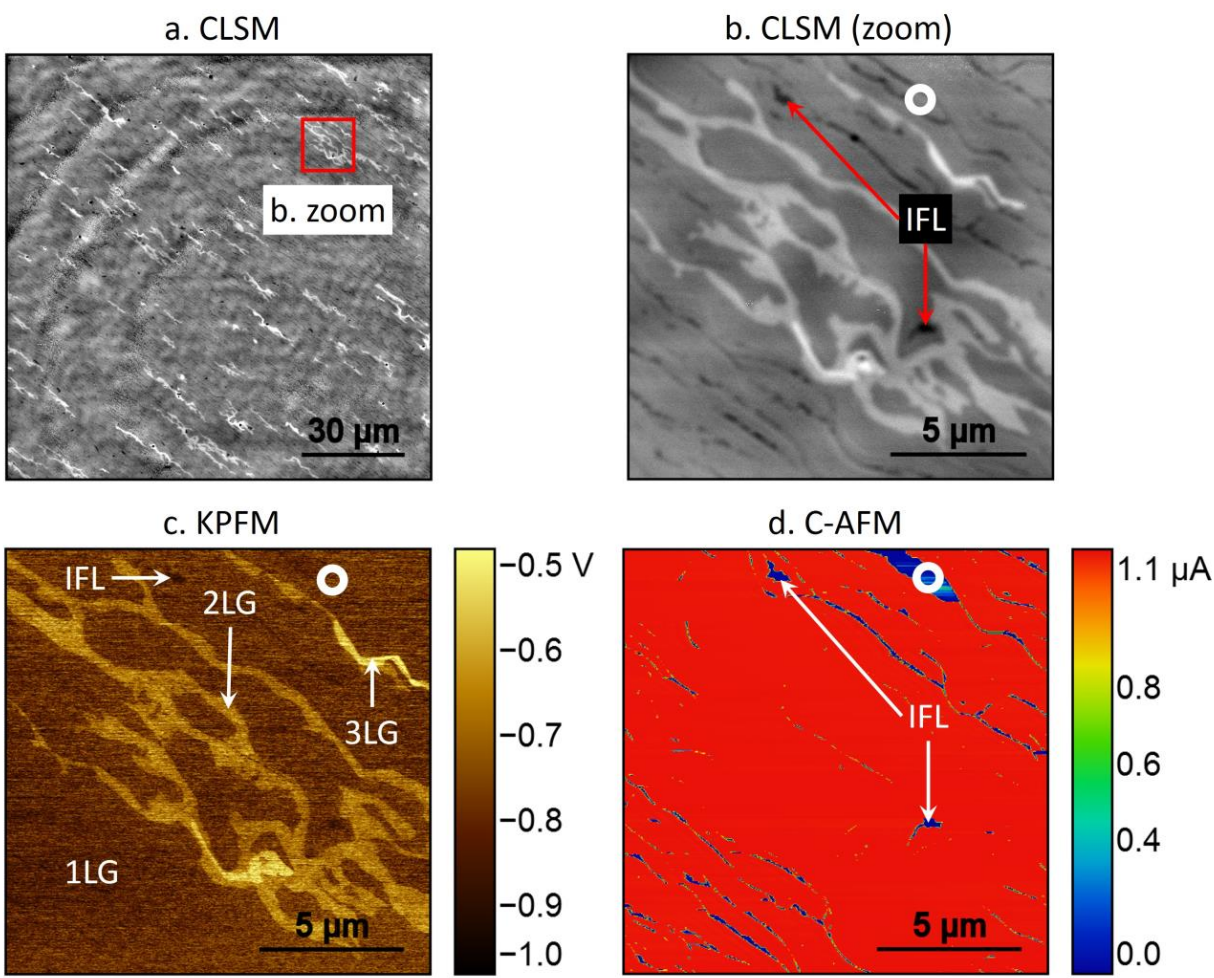

**Figure 3.** Large areas of monolayer EG with condensed bi- and tri-layer patches. CLSM image taken with (a) 100x and (b) 100x objectives with 8x digital zoom-in for the area indicated in (a). (c) Surface potential map of the same area as in (b) obtained by KPFM. (d) Current map of the same area as in (b) obtained by C-AFM.

## 3. Insusceptibility to surface charging

In Fig. 4, we further compared the CLSM images to SEM images for the two types of samples that have been shown in Figs. 1, 2 and 3. The SEM images were obtained in a vacuum chamber using the in-lens detector to capture backscattered electrons with beam energy less than 1 kV. Figures 4a and 4b are the CLSM and SEM images, respectively, of the same region from a graphene nanoribbon sample. The graphene regions appear brighter in the CLSM image (Figs. 4a and 4c), whereas the graphene appears darker in the SEM images (Figs. 4b and 4d). The inverted contrast can be explained by the higher work function of the graphene nanoribbons than that of the IFL or SiC, which leads to a stronger suppression of the backscattered electrons from the sample surface and therefore lower electron intensity sensed by the in-lens detector. Initially this appears to be inconsistent to the KPFM image (Fig. 3d), which suggests thicker layers of graphene have a lower work function, as the work function of the sample is given by $\Phi_{sample} = \Phi_{probe} - e\Delta V_{CPD}$, where $\Phi_{probe}$ is the work function of the probe[31]. However, as the KPFM image is obtained in ambient air (vs. vacuum for SEM), the atmospheric doping is likely to have a significant effect on the work function of graphene. If KPFM had been performed in vacuum as with SEM, the work function would be consistent, as demonstrated by Panchal et al.[39], where an inversion in the surface potential contrast was observed when placing the sample in vacuum.

Similarly, the SEM image of the overgrown sample (Fig. 4d) also has an inverted contrast compared to the corresponding CLSM image in Fig. 4c, since the work function of the multilayer patches is higher than the monolayer graphene, as indicated by the surface potential map obtained from KPFM in Fig. 3d. The lower resolution of Fig. 4d compared to Fig. 4b is most likely due to the screening effect from the continuous, conducting 1LG layer on top of the thicker layers. During the SEM scanning, the graphene surface becomes heavily charged and is also exposed to hydrocarbon contamination, both causing deterioration of the image resolution. On the contrary, no degradation of the CLSM image is observed *after* surface charging by electron beam, i.e., ***after*** SEM imaging.

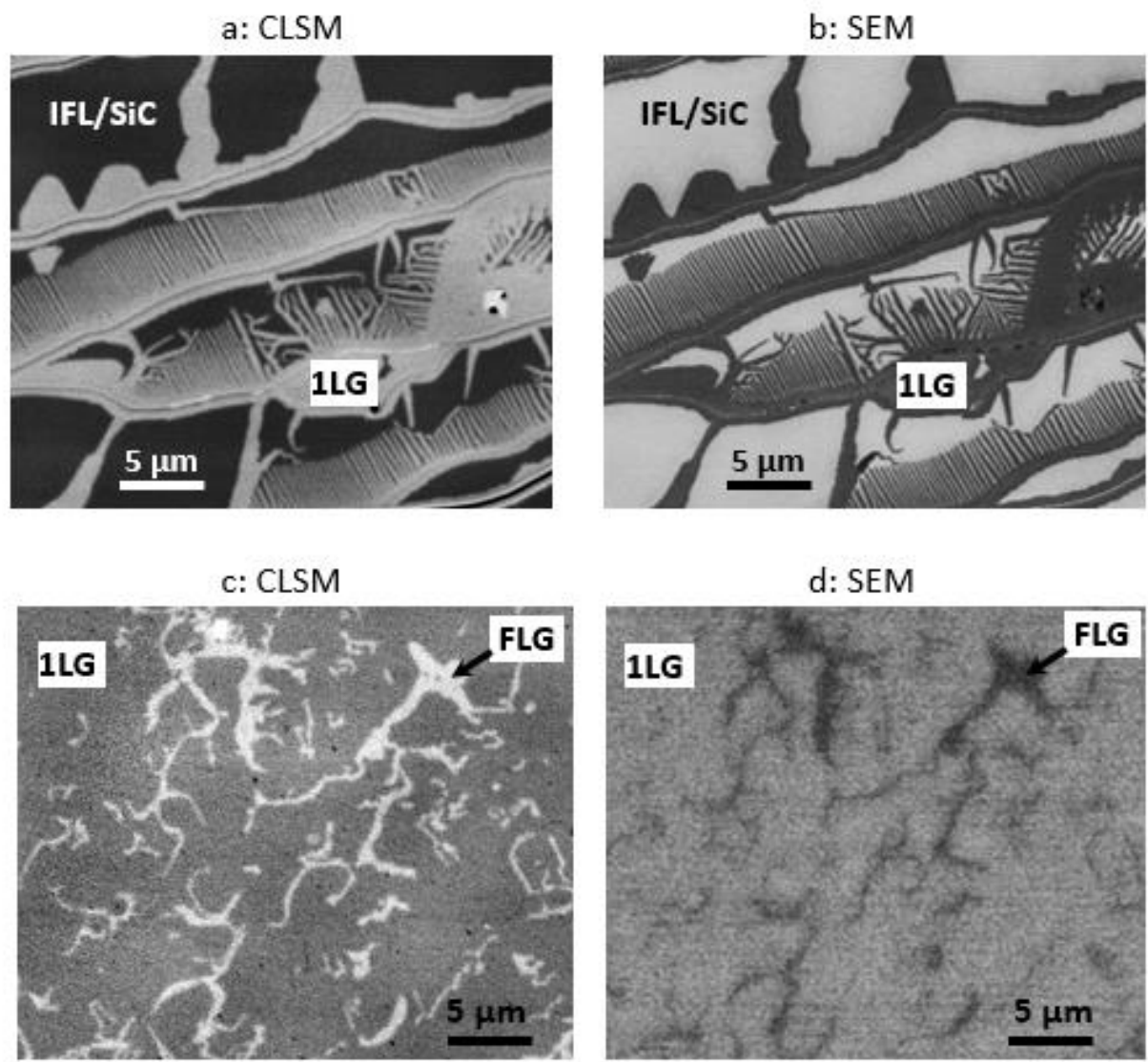

**Figure 4.** CLSM compared to SEM for EG on SiC. (a) CLSM image of the graphene ribbons on SiC. (b) SEM image of the same region as in (a). (c) CLSM image of multilayer graphene with uniform monolayer background. (d) SEM image of the same region as in (c).

**DISCUSSION**

In this paper, we have demonstrated fast and nondestructive characterization of epitaxial graphene by confocal laser scanning microscopy, which produces optical intensity images in ambient air, without any prior sample preparation, showing distinct features for incomplete, continuous, and overgrown graphene on SiC substrate. By analyzing the edge spread function, we estimate that the lateral resolution of our CLSM EG images is approximately 150 nm (Fig. S2 and S3), though it may be somewhat higher or lower under various imaging conditions, instrument settings and reflectivity of the sample. The measured reflected intensity from 1LG is $\approx$ 3% higher than that from adjacent IFL regions and the reflected intensity from 2LG is $\approx$ 2% higher than that from the 1LG (see note in Supplementary Information). Through the correlation to the results from Raman spectroscopy, SEM, and scanning probe microscope including AFM, C-AFM, KPFM, the CLSM images reveal that epitaxial graphene starts to develop from the edge of SiC terraces as parallel graphene nanoribbons. It then merges into a continuous, uniform monolayer graphene under proper processing conditions.

Micrometer-sized bilayer and few layer graphene patches are found as high contrast region in the CLSM images of overgrown samples. Compared to the supplementary methods used in this paper, CLSM not only has a much higher throughput for detecting such regions, it is also insusceptible to surface contamination or surface charging, which will strongly affect the resolution and even the contrast of KPFM or SEM images. Although Raman spectroscope also has advantages as an optical non-intrusive method, it suffers a much lower lateral resolution.

We propose that high resolution CLSM images can provide inspection of wafer-scale EG, selection of material and locations for more efficient fabrication (Fig. S5 and S6), as well as analysis of device quality and failure modes (Fig. S7). CLSM will be particular valuable for characterization of 2D materials, which have atomic thickness and are susceptible to surface contamination or surface charging, but have different reflectivity.

**METHODS**

The confocal laser scanning microscopy was performed using an Olympus LEXT OLS4100 system [see Notes] fitted with 5×, 10×, 20×, 50× and 100× objectives (numerical apertures: 0.15, 0.30, 0.60, 0.95 and 0.95, respectively) and with further 1× to 8× optical zoom. This enables the CLSM to image areas with field of view ranging from 2,560 μm to 16 μm, which translates to total magnification range from 108× to 17,280×. The system employs a 405 nm wavelength violet semiconductor laser, which is scanned in the X-Y directions by an electromagnetic MEMS scanner and a high-precision galvano mirror, and a photomultiplier to capture the reflected light and generate images up to 4096 × 4096 pixels with horizontal spatial resolution ≈ 150 nm (supplementary). The confocal optical setup only allows the reflected light that is in-focus to pass through the circular confocal pinhole, thus eliminating flare from out-of-focus regions, but resulting in a very shallow depth of field. To increase the focus resolving capability, a series of images along Z-axis are taken around the median focus height, with separations as small as 60 nm. For each pixel, an ideal *Intensity-Z* curve is calculated to fit the intensities in these images and extract the maximum value, which in turn is used to create a final 2D intensity image. The system is operated in ambient air and does not require any sample preparation for clean samples.

Tapping-mode atomic force microscopy (AFM) was performed in air using a Bruker Dimension FastScan SPM.[see Notes] In this mode, the probe tip oscillates at its fundamental resonance ($f_0$) and a feedback loop tracks the surface of the sample by adjusting the Z-piezo height to maintain a constant amplitude of the cantilever oscillation. The phase of the cantilever oscillation is also compared to the sine wave driving the cantilever oscillation, and thus, AFM achieves simultaneous mapping of the topography and tapping phase which is a measure of the energy dissipation between the probe and sample, thus encompassing variations in adhesion, composition, friction, viscoelasticity and other mechanical properties of the sample[38].

Conductive atomic force microscopy was performed using a Bruker Dimension Icon SPM [see Notes] by raster scanning a Pt/Ir coated probe across the sample surface. The C-AFM scans were performed with 250 mV bias voltage applied to the sample and the resulting current flowing through the probe at each pixel of the scan area was measured by a current amplifier. Epitaxial graphene's high electrical conductivity and good adhesion allow precise mapping of one layer graphene (1LG) or thicker few-layer graphene (FLG) structures by C-AFM unless they are isolated by non-conducting SiC or interfacial layer (IFL) carbon.

Kelvin probe force microscopy (KPFM) was performed by means of frequency modulation (FM) using a Bruker Dimension Icon SPM [see Notes]. During FM-KPFM, the surface of the sample is tracked and measured using the AFM feedback method described above. Additionally, a low frequency ($f_{mod}$), AC voltage ($V_{AC}$) is applied to the electrically conductive probe, which shifts the cantilever resonance due to the electrostatic attraction/repulsion and thus produces side lobes at $f_0 \pm f_{mod}$. When the FM-KPFM feedback loop applies an additional DC voltage to the probe ($V_{DC}$), the amplitude of the side lobes is proportional to the difference between $V_{DC}$ and the surface potential of the sample (also referred to as the contact potential difference, $V_{CPD}$). The surface potential is determined by the $V_{DC}$ minimizing the side lobes, i.e., when potential of the probe is equal to the potential of the sample. The surface potential map is obtained by recording $V_{DC}$ pixel by pixel. The surface potential values of the sample can be converted to a work function using, $\Phi_{sample} = \Phi_{probe} - e\Delta V_{CPD}$, provided the work function of the probe ($\Phi_{probe}$) is known. For further details see Ref. [39].

Raman spectra were acquired under ambient conditions with a 514.5 nm (2.41 eV) excitation (Renishaw InVia) [see Notes] which is focused to an approximately 1 μm spot on the samples through a 50× objective. Raman mapping measurements were performed by raster scanning an area of 28 μm by 20 μm with a step size of 2 μm and collecting the Raman G'(2D) peak region with an exposure time of ≈ 10 s for each point. Raman maps were generated by integrating the area of the Raman G'(2D) peak collected at each point.

**Acknowledgement**

A.F.R would like to thank the National Research Council's Research Associateship Program for the opportunity. The work of C.-I.L at NIST was made possible by arrangement with Prof. C.-T. Liang of National Taiwan University. We would like to thank Dr. Darwin Reyes-Hernandez for beneficial discussion about confocal microscope.

**Author Contributions**

V.P., Y.Y. and R.E.E. conceived and designed the experiments. V.P., Y.Y., G.C., J.H., C.-I.L., and R.E.E performed the experiments. Y.Y. and R.E.E. produced the samples. Y.Y. and C.-I.L.

fabricated the devices. Y.Y. and V.P. performed image processing and data analysis. The manuscript was written through contributions of all authors.

**Competing financial interest**

The authors declare no competing financial interest.

**Materials & Correspondence**

*Yanfei Yang, Email: yanfei.yang@nist.gov

*Vishal Panchal, Email: vishal.panchal@npl.co.uk

*Randolph E. Elmquist, Email: randolph.elmquist@nist.gov

**Additional Information**

**Supplementary Information** is available and includes (1) Evolvement of the optical contrast features from incomplete EG to continuous EG; (2) Estimation of lateral resolution for CLSM image; (3) Large area EG characterization by CLSM image stitching; (4) Notes on the reflected intensity from IFL, 1LG and 2LG; (5) Device inspection by CLSM.

**Funding Sources:** Work done by Y.Y. was supported by federal grant #70NANB12H185. Work done by V.P. at NIST and NPL was supported by federal grant and EC grant Graphene Flagship (No. CNECT-ICT-604391) respectively.

**Notes:** Commercial equipment, instruments, and materials are identified in this paper in order to specify the experimental procedure adequately. Such identification is not intended to imply recommendation or endorsement by the National Institute of Standards and Technology or the United States government, nor is it intended to imply that the materials or equipment identified are necessarily the best available for the purpose. The authors declare no competing financial interest.

# Supplementary Information:

# Rapid characterization of wafer-scale 2D material: Epitaxial graphene and graphene nanoribbons on SiC

Vishal Panchal[*1,2], Yanfei Yang[*1,3], Guangjun Cheng[1], Jiuning Hu[1], Chieh-I Liu[1,4], Albert F. Rigosi[1], Christos Melios[2,5], Olga Kazakova[2], Angela R. Hight Walker[1], David B. Newell[1], and Randolph E. Elmquist[*1]

[1] National Institute of Standards and Technology, Gaithersburg, MD 20899, USA

[2] National Physical Laboratory, Hampton Road, Teddington, TW11 0LW, UK

[3] Joint Quantum Institute, University of Maryland, College Park, MD 20742, USA

[4] National Taiwan University, Taipei, 10617, Taiwan

[5] Advanced Technology Institute, University of Surrey, Guildford, Surrey, GU2 7XH, UK

**Contents**

**1. Evolvement of the optical contrast features from incomplete EG to continuous EG**

Figure S1(a) is a reflective optical image of a FTG sample obtained by Nikon Eclipse L200N with a 50 x objective, showing incomplete single layer graphene (1LG) coverage. The graphene nanoribbons merged into continuous graphene in a succeeding growth, as shown in Fig. S1b. The conspicuous contrast from the interfacial layer regions (the darker contrast in Fig. S1a) disappeared in Fig. S1b. Instead, only narrow lines of higher brightness are seen after the second growth along the step edges, indicating few layer graphene, as confirmed by Raman spectroscopy. Figure S1c and S1d are cropped from Fig. S1a and S1b, respectively, showing the same region where a Raman map (Fig. S1e) has been generated after the second growth. The spectrum from a spot on the terrace, marked by a green circle in Fig. S1e, show a symmetric G'(2D) peak (the green curve in Fig. S1f) that can be fit by a single Lorentzian (the black dashed line in Fig. S1f), confirming the existence of 1LG. The spectrum from the spot at the step edge, marked by a red circle in Fig. S1e, shows a much wider asymmetric G'(2D) peak (the red curve in Fig. S1f), indicating few layer graphene.

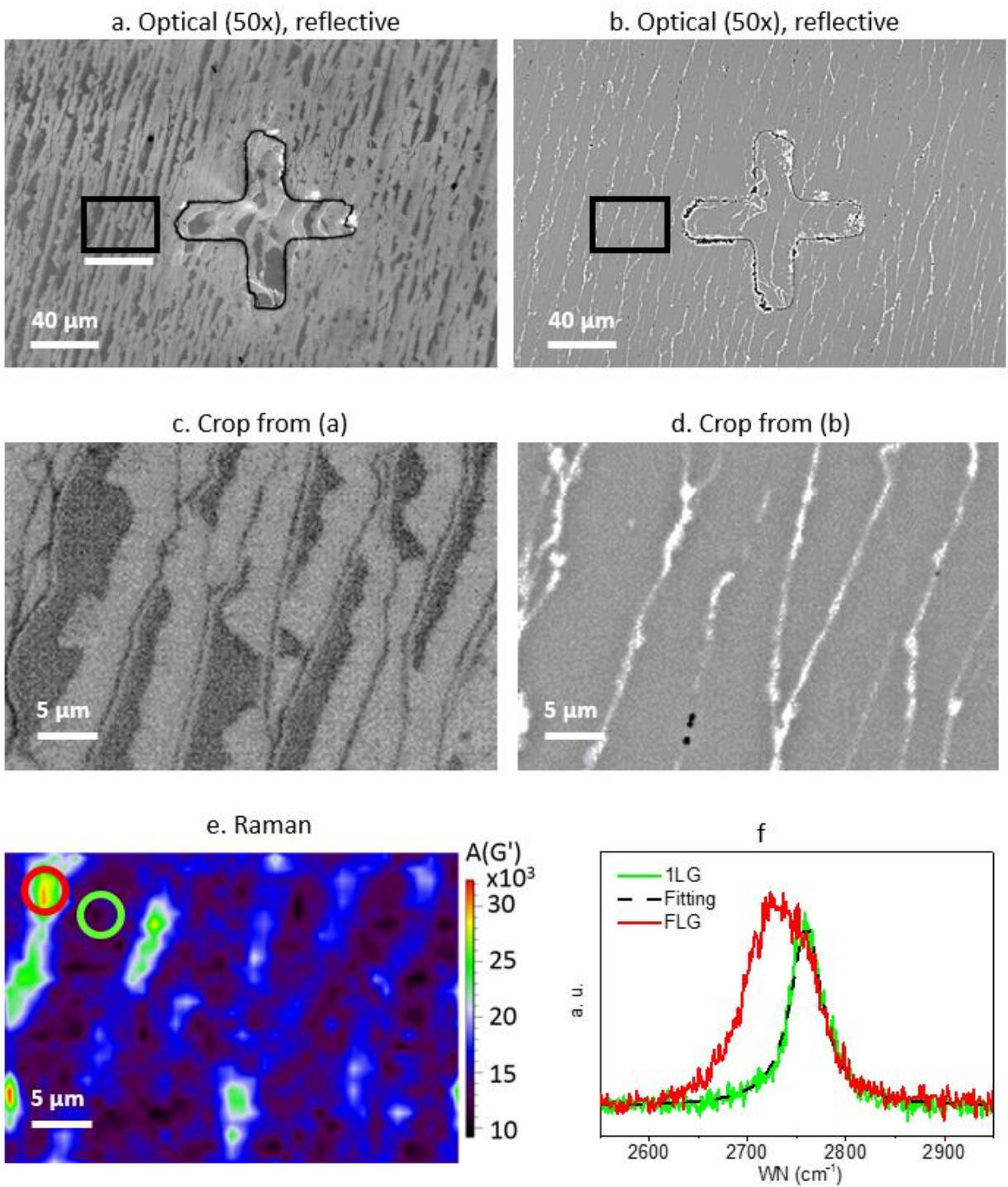

**Figure S1.** (a) Conventional reflective optical image of a FTG sample with partial graphene coverage. (b) Conventional reflective optical image of the same sample after a succeeding growth showing continuous background with narrow lines of higher brightness. (c) (d) Cropped images of the region marked by the black boxes in (a), (b) respectively. (e) Raman map of the same region as in (d) after the second growth. (f) Raman spectra from the spots marked by green and blue circles in (e), respectively. The black dot line is the Lorentzian fitting of the green curve.

## 2. Estimation of lateral resolution for CLSM image

Figure S2 shows the analysis of the same CLSM image as in Fig. 4a by a public image processing program ImageJ. The red curve in Fig. S2b is the averaged profile crossing the graphene nanoribbons marked by the red rectangular box, compared to the averaged profile (blue curve) from the same region of the SEM image in Fig. 4b. For the nineteen dips clearly seen from the SEM profile, seventeen corresponding peaks (counted from right) can be distinguished from the CLSM profile for the graphene nanoribbons with width varying approximately from 120 nm to 230 nm.

We further estimated the lateral resolution of the CLSM image in Fig. S2a by analyzing the edge spread function (ESF). The blue points in Fig. S2c and S2d are the averaged profile across the edges marked by two green rectangular boxes in Fig. S2a. An integrated Gauss function is used to fit the averaged profile, with the high plateau on the left defined as 100% brightness and the low plateau on the right defined as 0% brightness. The lateral resolution is estimated by calculating the edge width between two reference points with 20% and 80% of brightness (Fit 20/80). The inset label in Fig. S2c and S4d shows that the lateral resolution is approximately 149 nm and 161 nm for the left and right edges marked by the green boxes in Fig. S2a, respectively.

The lateral resolution will be affected by factors such as the contrast level and materials, and therefore varies from sample to sample, or from region to region on the same sample. Figure S3 shows the CLSM image taken with a 50x objective and 8x digital zoom. The lateral resolution values estimated from the "fit 20/80" of ESF for the denoted edges are listed in the table on the right, varying from approximately 97.3 nm to 185.8 nm. Based on analysis of more than 10 CLSM images from different samples, we estimate that the lateral resolution of our CLSM images is approximately 150 nm.

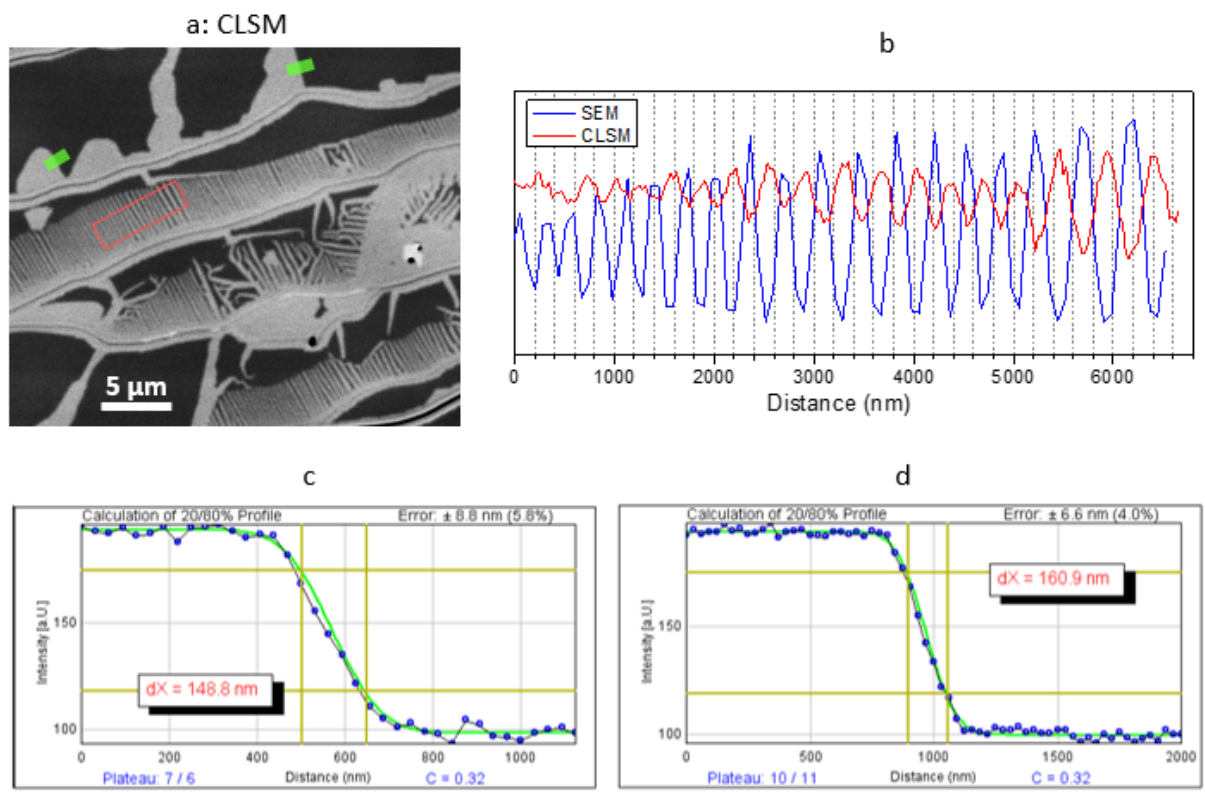

**Figure S2.** Analysis of the lateral resolution of the CLSM EG images by ImageJ. (a) CLSM of incomplete graphene on SiC (same image in Fig. 4a). (b) averaged profile for the red rectangular box in (a). (c) Gauss simulation (green line) of the edge indicated by the green rectangular in (a) on the left. (d) Gauss simulation (green line) of the contrast edge indicated by the green rectangular in (d) on the right.

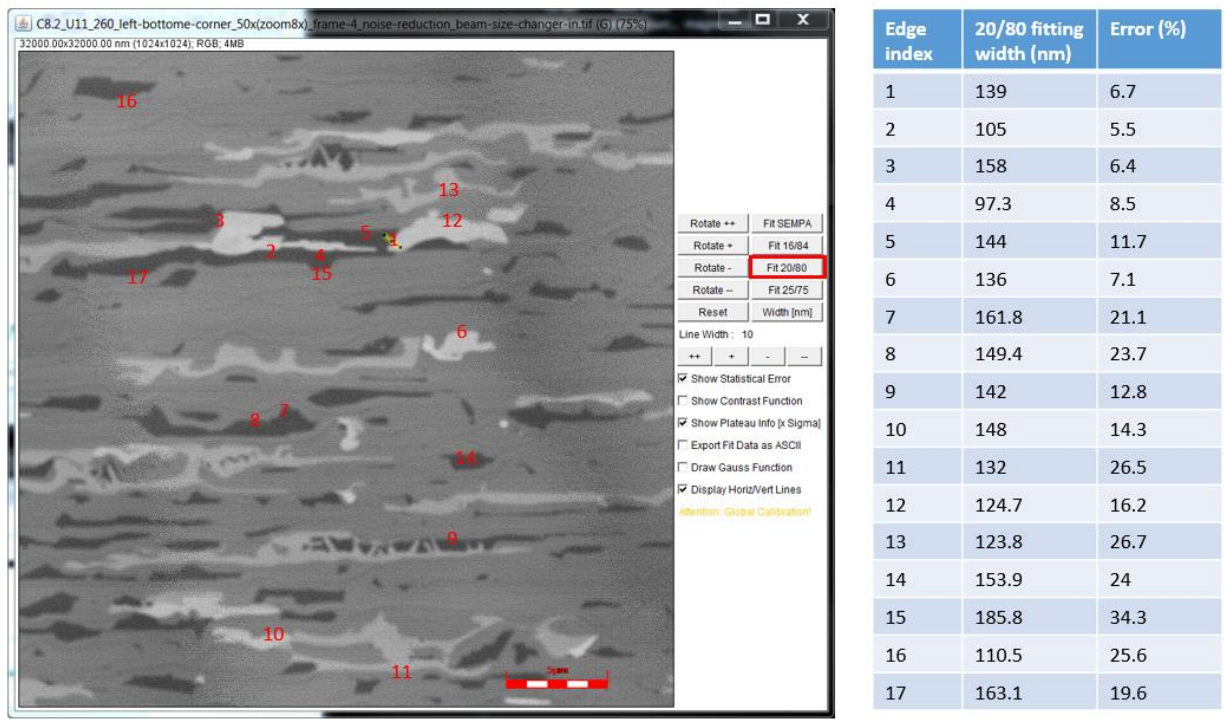

| Edge index | 20/80 fitting width (nm) | Error (%) |
|---|---|---|
| 1 | 139 | 6.7 |
| 2 | 105 | 5.5 |
| 3 | 158 | 6.4 |
| 4 | 97.3 | 8.5 |
| 5 | 144 | 11.7 |
| 6 | 136 | 7.1 |
| 7 | 161.8 | 21.1 |
| 8 | 149.4 | 23.7 |
| 9 | 142 | 12.8 |
| 10 | 148 | 14.3 |
| 11 | 132 | 26.5 |
| 12 | 124.7 | 16.2 |
| 13 | 123.8 | 26.7 |
| 14 | 153.9 | 24 |
| 15 | 185.8 | 34.3 |
| 16 | 110.5 | 25.6 |
| 17 | 163.1 | 19.6 |

**Figure S3.** Left: CLSM image of a EG sample covered by dominant single layer graphene (1LG). The lowest brightness indicates IFL region. Higher brightness corresponds to thicker graphene layer. Right: A table with edge width calculated from "Fit 20/80" algorithm.

### 3. Notes on the reflected intensity from IFL, 1LG and 2LG

We have found that the sharpness level (Fig. S4) in the advanced settings for the CLSM will strongly affect the reflected intensity due to the backstage algorithm. As suggested by the Olympus specialist, we turned off the contrast and sharpness when we estimate the ratio of reflected intensity from IFL, 1LG and 2LG. We have universally observed that the reflected intensity from 1LG is $\approx$ 3 % higher than that from IFL region, and the reflected intensity from 2LG is $\approx$ 2 % higher than that from 2LG.

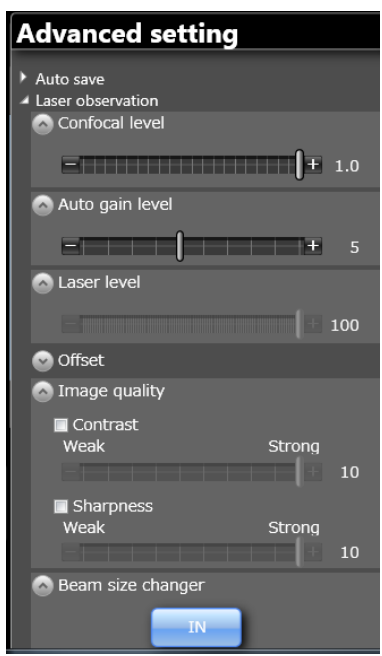

**Figure S4.** CLSM advanced setting used for estimation of the change of reflected intensity from IFL, 1LG and 2LG.

### 4. Large area EG characterization by CLSM image stitching

Since the graphene nanoribbons as well as the 2LG and FLG patches are usually submicron sized, a single CLSM scan by objective 20X and higher magnification cannot distinguish such features properly and are not suitable for the characterization of EG region larger than hundreds of micrometers. Wafer-scale EG can be characterized by stitching arrays of CLSM images scanned by 50x or 100x objective as shown in Fig. S5 and Fig. S6.

Figure S5a shows a high resolution CLSM image (produced from 64 CLSM scans by digital stitching) of a homogeneous monolayer graphene area (463 μm by 463 μm) that includes less than 1% of multilayer graphene (the irregular brighter patches). Figures S5b,c,d are the three zoomed-in grid CLSM images for locations marked by red boxes 1,2, and 3 in Fig. S5a. Hall bar devices of 400 μm width fabricated from such graphene can maintain the quantum Hall effect with precise metrological accuracy up to 4 K.

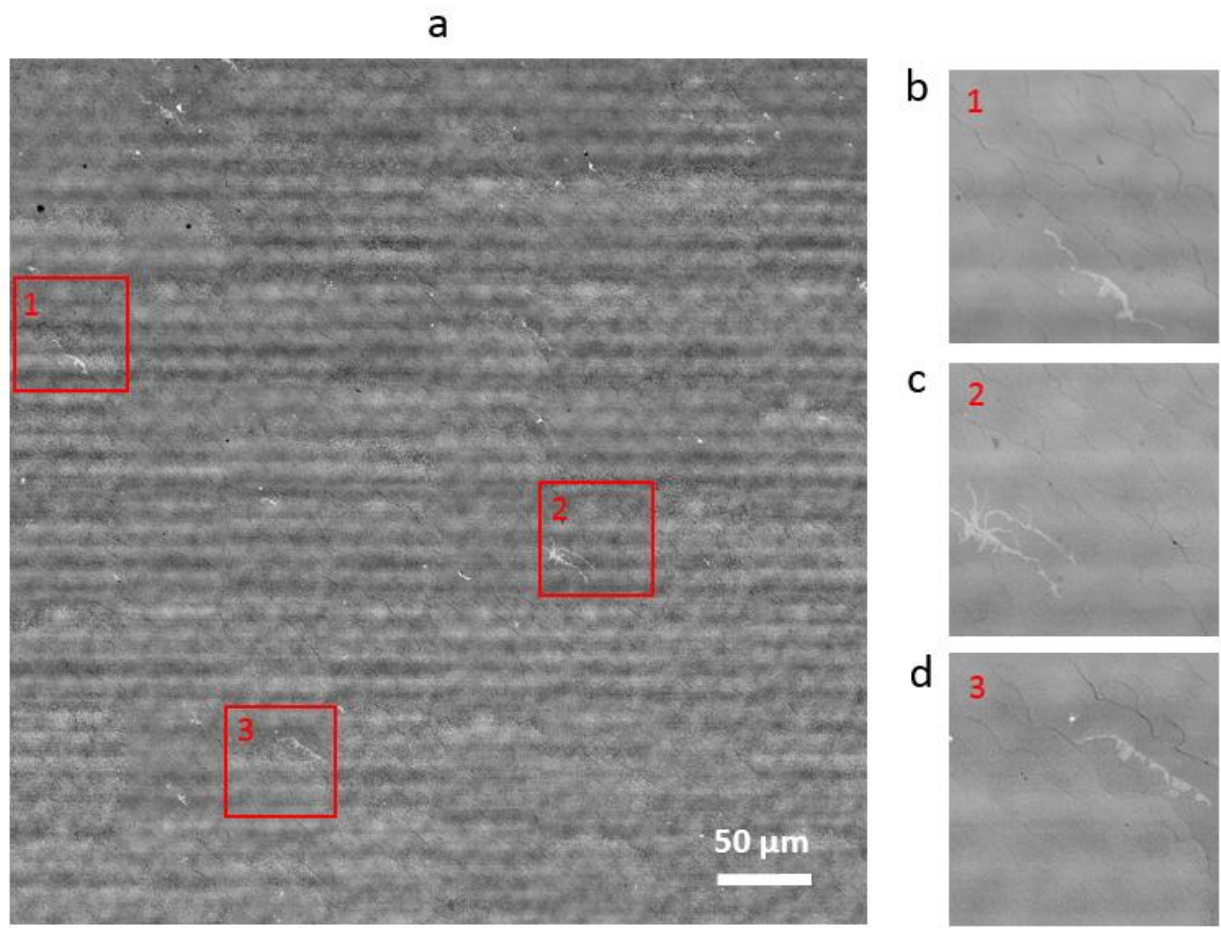

**Figure S5.** (a) Stitched CLSM image of a highly uniform area of monolayer graphene with only few bilayer patches as shown in the right panels. (b) Zoom-in of the region indicated by the red box 1 in (a). (c) Zoom-in of the region indicated by the red box 2 in (a). (d) Zoom-in of the region indicated by the red box 3 in (a).

Fig. S6 shows a composite image produced in ≈ 20 minutes from 16 CSLM scans by digital stitching. The black strip that appears in the lower region of this image is the edge of the sample. The fiducial mark (V20) is used for sample identification, and is etched into the SiC before EG is grown. FTG growth usually produce very thick graphene layers (Ref. 5) close to the edge of the sample (region 3 with much higher brightness in Fig. S6). About few hundreds of micrometers away from the edge, bilayer and few layer patches decrease dramatically in region 2. Continuous

EG with less than 1% of bilayer or few layer patches in region 1 is suitable for fabrication of quantum Hall resistance standards (Ref. 5).

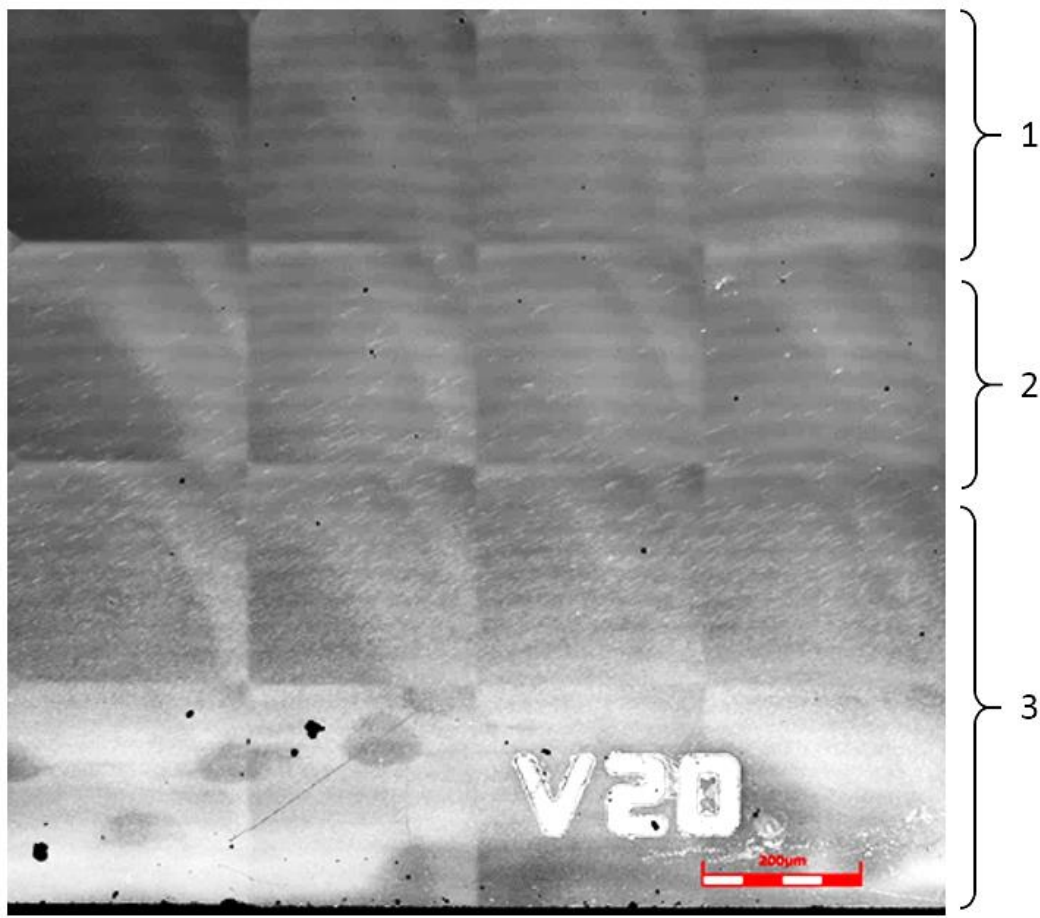

**Figure S6.** Stitched CLSM image of an area near the edge of a primarily monolayer EG sample, where thicker graphene patches (bright contrast) can only be seen near the edge. Region 1 is covered by uniform and continuous 1LG. Region 2 shows increasing bilayer and few layer patches (with higher brightness). Region 3 is covered by very thick graphene.

## 5. Device inspection by CLSM

Charge carrier mobility of graphene is an important electronic property that is usually measured through Hall effect. However, the mobility of epitaxial graphene is strongly affected by its carrier density. To compare the quality of two graphene devices, one needs to compare the curves of mobility as a function of carrier density obtained at low temperature, as shown in Figure S3. Here we correlate the mobility characteristic curves to the CLSM images of corresponding devices. The CLSM image (left inset in Fig. S3) of the high mobility device (red data in Fig. S3) shows almost complete graphene coverage with less than 10% of bilayer or interfacial layer inclusions. The CLSM image (right inset in Fig. S3) of the low mobility device (black data in Fig. S3) shows large portion of interfacial layer.

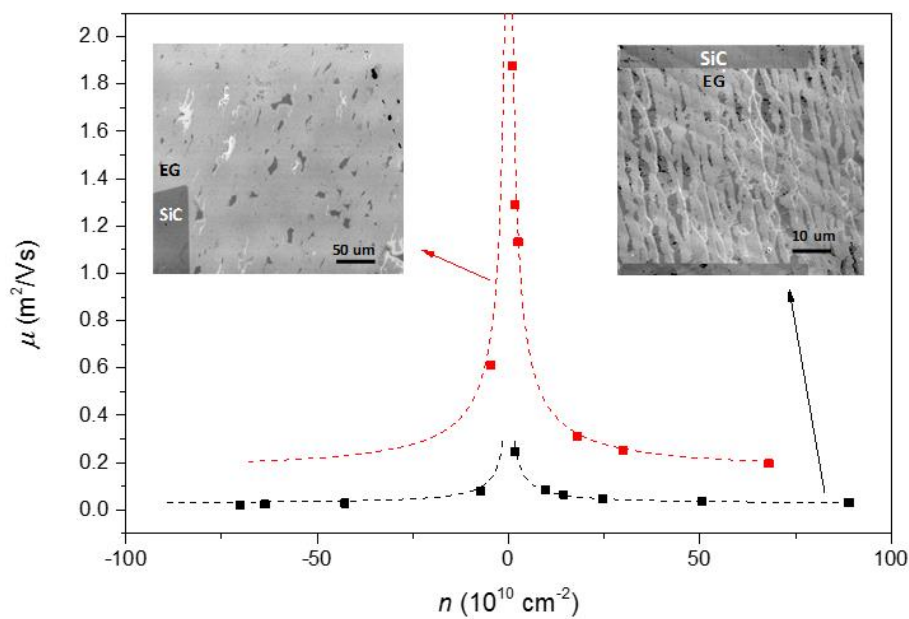

**Figure S7.** Mobility of EG devices as a function of carrier density. The overall mobility of a more uniform sample (data shown in red, CLSM image in left inset) is much higher than that of another device (data shown in black, CLSM image in right inset) made from graphene area with incomplete graphene and nanoribbons.